**Title:** Wide Area Surface Dosimetry with Conformal Scintillator Array for External Beam Radiotherapy

Roman Vasyltsiv[1], Allison L. Matous[2], Natasha Mulenga[1], Megan A. Clark[1], Brian W. Pogue[1], David J. Gladstone[1,2], Lesley A. Jarvis[2], Petr Bruza[1]

[1]Thayer School of Engineering, Dartmouth College, Hanover, NH

[2]Department of Radiation Oncology and Applied Sciences, Geisel School of Medicine, Dartmouth College, Hanover, NH

**Abstract:**

**Background:** In vivo dosimetry is essential for treatment verification in modern radiotherapy, but existing techniques have significant limitations in their logistics of use related to time, accuracy of placement, temporal resolution, and non-uniform anatomy. Optical scintillation imaging dosimetry has shown potential to address several of these limitations, and the translation to conventional photon external beam radiotherapy was examined with a novel wide area imaging and sensing technique.

**Purpose:** This study characterized a conformable scintillator array, for imaging real-time surface dose delivery over a wide curved surface area in conventional photon external beam radiotherapy. The variability was evaluated for dose, angular deviation, repetition rate, and field geometry while assessing its performance in field edge detection and measurement reproducibility using an anthropomorphic torso phantom.

**Methods:** The modular scintillator array consisting of 101 hexagonal scintillating elements was used together with a dual-camera stereovision system for 3D localization and a clinical Cherenkov imaging system for optical scintillation detection. The water equivalent thickness, dose linearity (10-800 MU), repetition rate stability (60-600 MU/min), and angular dependence of the scintillator array were characterized. Field edge detection was evaluated against radiochromic film through gamma analysis. System reproducibility was assessed through five consecutive deliveries simulating contralateral breast monitoring during whole breast treatment with a tangent beam.

**Results:** The scintillator array demonstrated high dose linearity ($R^2 = 0.999$) across the full tested range, with minimal beam perturbation (1.21 mm water equivalent thickness) and consistent response within 5% across all available LINAC repetition rates. Dose-normalized scintillation varied within 5% for clinically relevant gantry angles (0-75°), while camera angle corrections successfully compensated for deviation from Lambertian emission. Field edge detection showed high agreement with a 99.98% gamma pass rate (3%/3mm) compared to film. Inter-delivery reproducibility of ±1 cGy demonstrated robust system performance across multiple acquisitions.

**Conclusion:** The conformable scintillator array imaging system provides spatially resolved, dynamic surface dosimetry with minimal workflow impact. Its ability to generate continuous dose maps across complex anatomical surfaces while maintaining angular correction capabilities addresses key limitations of current in vivo dosimetry approaches. This technology shows promise for clinical integration, particularly for treatments where wide area surface dosimetry is necessary.

## 1. Introduction

External beam photon radiation therapy has undergone significant advancement in control over beam positioning, shaping, and delivery, leading to widespread adoption of advanced delivery management techniques. This progression has led to increased treatment plan complexity to achieve improved dose conformality to targets, and better normal tissue sparing while maintaining tumor control.[1–3] Achieving this increase in precision of dose delivery required a greater need for treatment accuracy emphasizing the importance of patient positioning, treatment setup, and verification. Verification of accurate delivery now requires high temporal and spatial resolution dose assessment in patient coordinates with direct reference to the treatment site.[4,5] Given the complexity of these requirements, the current standard of performing pre-treatment dosimetry quality assurance is insufficient as it cannot catch errors related to patient geometry or in beam delivery during the actual treatment. In vivo dosimetry (IVD) is therefore an attractive tool as it allows the dose delivery to be monitored locally, providing a reference and verification of the treatment following the delivery..[5–7] However, despite these advantages, achieving precise measurement remains challenging.

Conventional IVD techniques present significant drawbacks, especially when measuring dose over multi-field deliveries, wide areas of non-uniform anatomy, or steep dose gradients – all of which are common factors in modern treatments. Existing techniques generally involve either transit or on-patient dosimetry, with the latter being further subdivided into point dosimetry and film. Transit dosimetry primarily uses electronic portal imaging devices (EPIDs) to validate exit dose.[8] Though effective for temporal monitoring and validation of field structure, it offers limited ability to quantify absolute dose within the patient due to complex, model-dependent corrections required for dose back-projection.[5,8] Point dosimetry employs single spot dose reporters such as thermoluminescent dosimeters (TLDs), optically stimulated luminescent dosimeters (OSLDs), diodes, or MOSFETs to accumulate dose throughout the treatment duration.[5,9] In the case of TLDs and OSLDs, processing may take several days as the charge is stored locally and must be read out at a location with the necessary equipment. Importantly, point dosimetry also lacks reliable correlation to treatment plans due to placement without guided positioning or post-treatment localization, introducing a critical limitation in dose gradient regions where positioning errors can lead to dosimetric misinterpretation. Gafchromic film offers two-dimensional dosimetry but has limited application for in vivo monitoring due to its semi-rigid structure and flat readout requirement, restricting conformability to curved surfaces and potentially distorting surface dose profiles.[10] Additionally, film typically requires delayed readout, preventing intra-treatment or inter-field monitoring.

An emerging method of optical scintillation imaging dosimetry (SID) has shown promise in addressing many of the main drawbacks of existing IVD techniques.[11–14] SID leverages the linear response of scintillating materials with dose and dose rate together with in-room camera imaging to collect scintillator emission during the beam delivery and relate the intensity to dose. Several attempts to image scintillation on the body surface have been demonstrated to pioneer the technique, using scintillating sheets[15–17] and wearable fabric options[18,19]. However, flat scintillator sheet imaging cannot be translated in vivo given the inherent inability to conform to surface geometry. Fabric scintillator imaging is likewise limited by material homogeneity and variability in emission and has so far only been shown in qualitative applications, lacking dosimetric comparison to plan or reference dosimeter.[17,19] Furthermore both approaches lack the ability to identify and correct for the known intensity variation with increasing imaging angle[11,13,20] which is fundamentally required for optical 2D dosimetry in clinical applications. Prior work suggests that stereovision imaging can identify deformation vector fields,[21] but this has not yet been applied to 2D dose monitoring and correction.

Prior studies using solid single-point scintillator imaging with commercial Cherenkov imaging systems demonstrated feasibility for point dosimetry with additional field monitoring through simultaneous Cherenkov imaging.[13,22] Rigid scintillation sheets have shown feasibility for time-resolved full field quality assurance in photon and proton beams[14,23], confirming adequate spatial and temporal resolution for dose monitoring. Recently, a deformable scintillator array has been developed for ultra-high dose rate proton therapy surface dosimetry.[11] Constructed of rigid scintillating elements connected by a flexible mesh, this system monitors delivery with sub-millisecond resolution, while conforming directly to the target geometry and maintaining linear response with minimal water equivalent thickness. Implementation with dual-camera stereovision enables array localization within the treatment environment and real-time structure reconstruction. While this approach is promising, it was tested in UHDR PBS proton therapy conditions with known pencil beam structure and negligible build up effects. This study extends scintillator array dosimetry to external beam photon therapy, representing the first comprehensive assessment for translating this methodology to conventional radiotherapy. We evaluate array impact on beam delivery, scintillation response to angular variations, and geometric processing requirements. We demonstrate the feasibility of scintillator array dosimetry for field edge monitoring, and evaluate the application of this method using an anthropomorphic phantom. Specifically, this study is done in preparation for clinical translation to in vivo monitoring of contralateral breast dose during tangent breast irradiation given the high risk of secondary cancer development due to low dose spill from the ipsilateral breast treatment.[24] This application also serves as a uniquely challenging case for standard in vivo dosimetry as the contralateral dose spill is often in a high dose gradient region, demonstrating the requirement for 2D dose monitoring and high target conformity.

## 2. Methods

Scintillator array imaging has previously been evaluated as a surface dosimetry technique for UHDR PBS proton therapy. This approach was translated here to EBRT photon therapy by imaging with clinical optical camera hardware optimized for Cherenkov imaging. Additionally, photon therapy uses wide area field irradiation which differs from the known structure of PBS proton therapy, requiring an alternative approach to inter-element interpolation. Finally, angular impact of scintillator emission and beam impact of the array were evaluated to ensure reliable intensity to dose translation. The work herein addresses these concerns and evaluates scintillator array imaging as a surface dosimetry tool with a sample application of contralateral breast dose monitoring on an anthropomorphic torso phantom, depicting a common treatment application with non-uniform geometry and high dose gradients at the field edge.

### 2.1 System Setup and Overview

The experimental setup used to evaluate scintillator performance and feasibility of translation is shown in Figure 1a and consisted of a dual-camera stereovision system mounted adjacent to an FDA approved Cherenkov camera [BeamSite, DoseOptics, Lebanon, NH] and a scintillator array placed atop an anthropomorphic chest phantom. The scintillator array used in this work is shown in Figure 1b. The structure is expandable to a maximum size of 30x30cm but has been constrained to a 10x4cm region given the partial field edge dosimetry application and consists of 101 hexagonal elements of a blue-emitting plastic scintillator (Penn-Jersey X-Ray, Blue-800) with a 7mm inscribed diameter, a 0.5mm opaque surrounding wall, and a 0.5mm inter-element separation. The central element was removed to facilitate accurate visualization and placement over the target area and allow for a reference dosimeter (e.g. TLD) placement in future clinical translation. Camera setup details are shown in Figure 1c. The Cherenkov camera (17 frames/s, f/2, 50 mm) is part of a clinically installed system oriented toward the LINAC isocenter at a distance of 2.2 m and introduces up to a factor of 10x higher scintillation signal compared to background Cherenkov at f/2, allowing for scintillation imaging without SNR constraints. The stereovision dual-camera system (60 frames/s, f/4, 50 mm) was designed to mount adjacent to the Cherenkov camera, enabling data collection without disturbing treatment workflow or clinical monitoring. The stereovision modules also presented a significantly higher resolution (2448x2040 pixels) compared to the Cherenkov camera, which directly improved accuracy of stereovision localization by introducing finer sampling in the disparity-to-depth relationship. Camera separation was set at 14 mm based on of the 2.2m working distance and wide field-of-view lenses. An inward turn angle of $\theta = 3°$ was set to orient the target array at the center of both stereovision imaging views (shown in Figure 1d and 1e). Optical surface emission from a sample tangent field irradiation is shown (Figure f-g), leveraging the full field visualization of Cherenkov imaging while simultaneously providing quantitative reference in the region within the scintillator array.

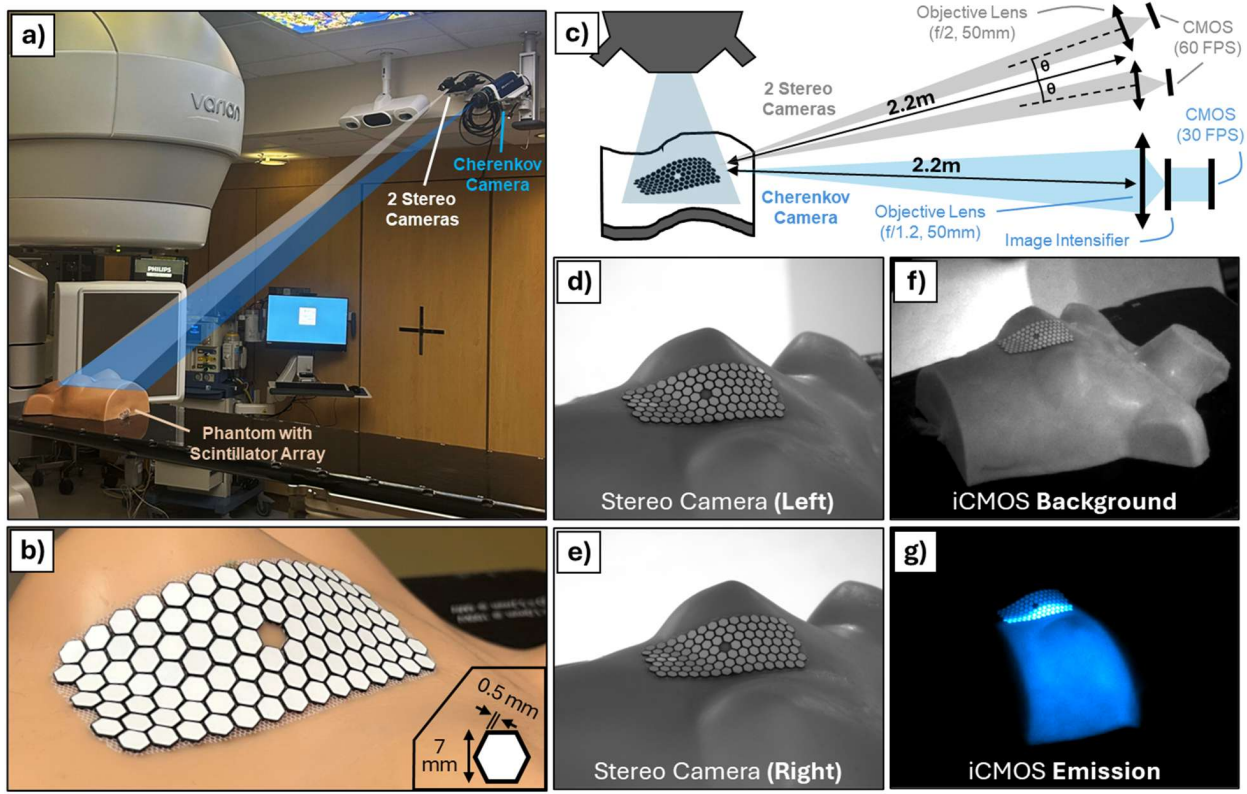

**Figure 1:** a) Sample experimental setup showing a dual-stereovision system mounted to the ceiling in a treatment bunker adjacent to a clinical Cherenkov camera. (b) A detailed view of the scintillator array is shown in a sample position on an anthropomorphic chest phantom. (c) A reference diagram of the imaging setup is highlighted and sample image outputs from the stereovision modules (d-e) and Cherenkov camera (f-g) are shown.

**2.2 Data Acquisition and Pre-Processing**

Optical emission from the scintillator array during irradiation was captured using the clinical Cherenkov BeamSite system. The intensified CMOS camera is gated to the LINAC pulse sequence, sequentially capturing frames of optical emission during beam delivery and subsequent background frames, each at 16 frames/s. Following the acquisition, the image stacks undergo darkfield correction, spatial and temporal filtering, and background subtraction to isolate the radiation induced optical signal (Figure 1g). The resulting background and Cherenkov images are saved as 32-bit tiff files. The Cherenkov images allow for delivery-gated capture of the scintillation signal and were used for downstream dose map conversion.

The dual-camera stereovision system was designed as a passive addition to the clinical Cherenkov camera, involving no interference or interaction with the clinical hardware. The stereovision cameras monitored the array position throughout the delivery at 60 FPS with simultaneous image capture (<2 ms delay). Images were saved as 8-bit files at a resolution of 2448 × 2048 pixels. A custom guided user interface (MATLAB 2022b) was developed to enable remote acquisition control during beam delivery. Prior to beam monitoring, the stereovision system underwent intrinsic and extrinsic calibration where a structured point board (4x11 points, 20 mm separation) was imaged in 20+ orientations by both cameras. Using the MATLAB Stereo Camera Calibrator application, extrinsic and intrinsic parameters were derived and saved. The calibration procedure had to be redone whenever the relative orientation of the cameras was adjusted.

To pre-process the Cherenkov imaging data, the background image was histogram equalized, and the hexagonal array was segmented using local contrast enhancement and adaptive thresholding. Given that the camera position and field of view remained the same, scintillator segmentation was constrained based on empirically derived expected size range of the hexagonal elements, enabling automatic removal of any background or noise that remained after thresholding. The resulting mask was used to scale the Cherenkov image and isolate the pixels that correspond to scintillator emission. The same procedure was then done to isolate the scintillator array profile in the stereovision images. Finally, the stereovision extrinsic and intrinsic parameters were used to rectify the stereovision image pair (Figure 2a) such that corresponding features in image 2 laid on the same row as their counterparts in image 1.

**2.3 3D Localization and Array Surface Fusion Using Stereovision**

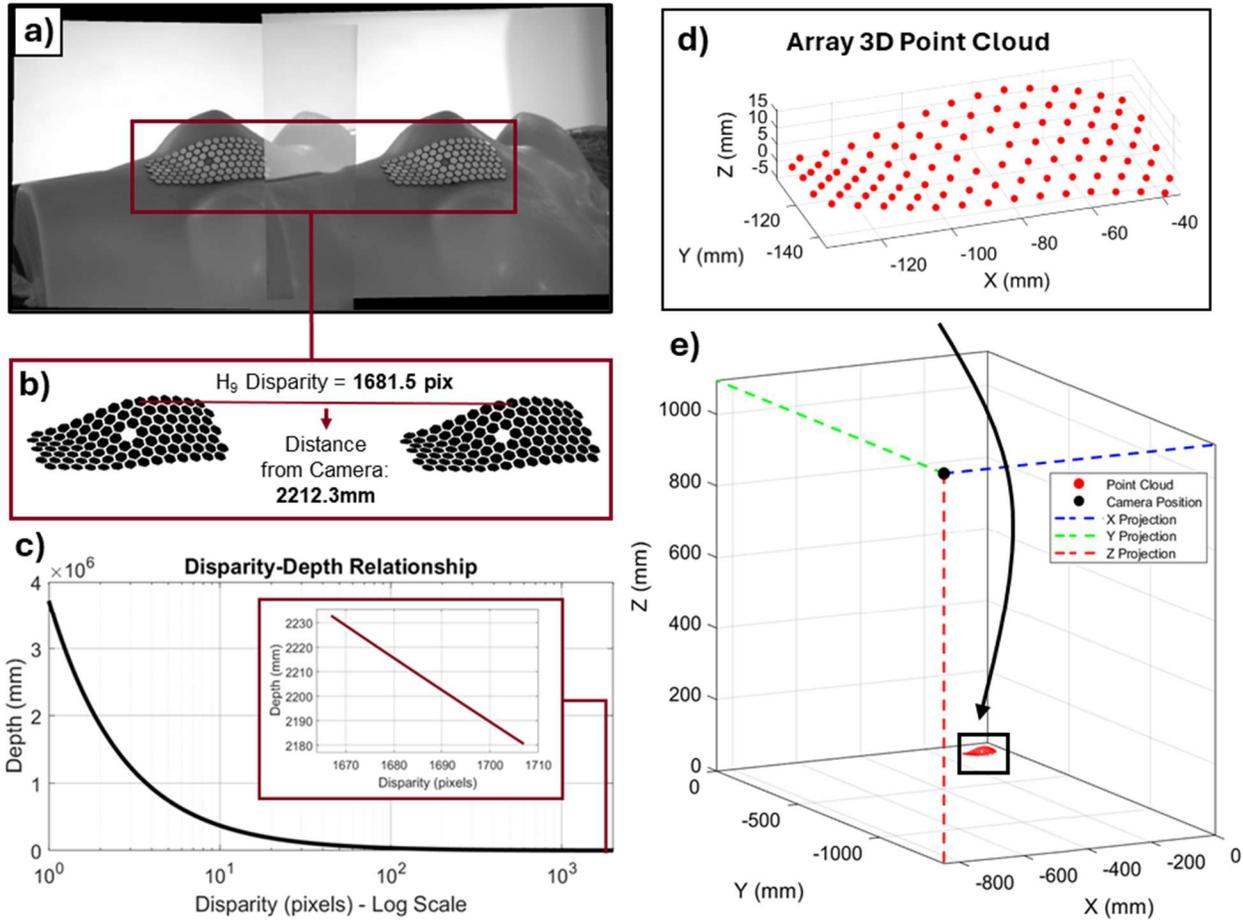

**Figure 2:** a) Shown is a sample rectified image pair, b) with a corresponding sample disparity definition using masked array profiles, c) and an analytically derived relationship between disparity and depth. d) A sample array point cloud derived from the stereovision relationship is shown and is oriented in e) the bunker environment with respect to the isocenter and the camera.

Ideally, an effective 2D surface dosimeter should not only conform to the target geometry, but also be localizable in the 3D room coordinate space for verification of placement and positioning. This is particularly important for scintillation-based systems as scintillators have been shown to exhibit angular dependencies in emission which must be corrected prior to relating intensity to dose.

The scintillator array dosimetry system uses stereovision to localize the position of each scintillating element in 3D and in room coordinates. Following image rectification of the grayscale (Figure 2a) and binarized array (Figure 2b) images, image disparity is calculated only for the centroid pairs between related hexagonal elements in the two images. Disparity is defined as the difference in horizontal position of the same point in the left and right images and directly relates to the relative depth of the given pixel in 3D coordinates, as shown in Figure 2c and translated by:

$$Z(m,n) = \frac{f \cdot B}{d(m,n)}$$

Where Z is the depth defined as the distance from the camera, f is the focal length of the imaging system, B is the distance between the stereovision camera modules, and d is the disparity for a given pixel. Importantly, the disparity definition is restricted only to the centroids of each hexagon, resulting in a 3D point cloud that consists only of the scintillating element positions, as shown in Figure 2d. The external stereo parameters are then used to orient the point cloud in room coordinates with respect to the isocenter (Figure 2e), relating the position of the array to that of known fixed reference points such as the camera and gantry.

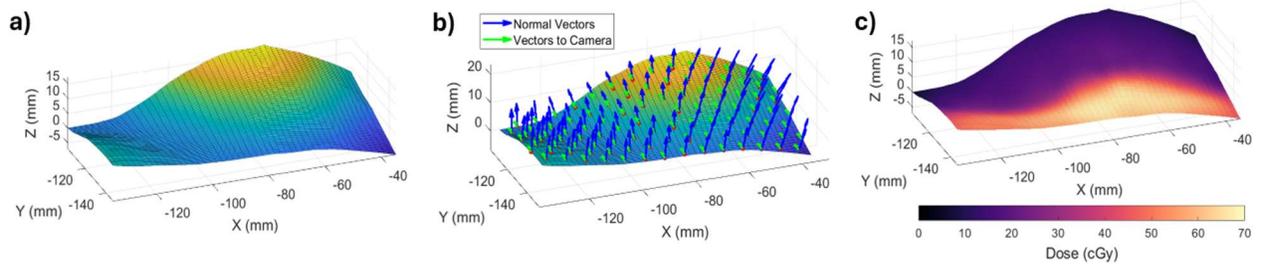

**Figure 3:** Figure showing the a) interpolated surface profile with b) an overlay of the calculated normal vectors for each scintillating element orientation and directional vectors toward the camera position, and a c) sample dose map following angular correction and interpolation.

The point cloud data is used to generate a continuous 3D surface (Figure 3a) through natural cubic spline interpolation in MATLAB, which minimizes overall curvature and closely reflects human anatomical contours. Normal vectors are calculated at each point on this surface and combined with directional vectors to the Cherenkov camera (Figure 3b) to determine the angular deviation for each scintillator element. This 3D surface profile serves as the reference framework for the dose map derived from scintillator emissions across all 101 elements. The proximity and orientation of stereo modules and Cherenkov camera allow for direct correlation between images based on the binary masks and similarity of the individual hexagonal element positions, enabling centroid mapping between hexagons in Cherenkov images and corresponding points in stereo images. Individual hexagons can be independently scaled to account for camera deviations, as each element was segmented and mapped to its 3D centroid. Finally, intensity values from each hexagonal scintillator were interpolated to adjacent neighbors using cubic spline interpolation, effectively filling gaps inherent to the fragmented array design and yielding a continuous surface dose map (Figure 3c).

**2.4 Dose Linearity and Repetition Rate Stability**

Scintillation emission as a function of monitor units (MU) was compared against Gafchromic film. The scintillator array was positioned at the center of a 10×10 cm field at isocenter on top of 5 cm of solid water. One end of the array was secured to the solid water to prevent motion between deliveries, while a sheet of gafchromic film was placed underneath the array in the same position for each delivery. The setup was irradiated with a 6 MV photon beam at 10, 50, 100, 300, 500, and 800 MU, ensuring capture of the known linear range for gafchromic film while also verifying response at low dose levels. Linearity with MU was evaluated for both the array and the film, with 1 MU defined as 1cGy at the depth of maximum dose for a 10x10cm field delivery.

Scintillation emission as a function of repetition rate was evaluated using the same setup as the dose linearity study. A 6MV photon beam was used to deliver 500 MU at varying repetition rates of 60, 100, 200, 400, and 600 MU/min. The integral scintillator output was normalized by film dose to account for any variability in delivery and subsequently analyzed for constancy across the different repetition rates.

## 2.5 Angular Dependence

### 2.5.1 Dependence on Beam to Scintillator Angle

To characterize the dependence of scintillator emission on gantry angle, the array was placed onto a flat 5 cm block of solid water and aligned to the isocenter. A sheet of Gafchromic film was placed underneath the array and a 15×15 cm$^2$ 6 MV field was used to deliver 500 MU. The resulting scintillator emission was recorded and the average response along the central portion of the array was normalized by the dose of the corresponding film. The procedure was repeated at 10-degree intervals from 0 degrees (anterior-posterior) to 180 degrees (posterior-anterior), replacing the film with each iteration.

### 2.5.2 Dependence on Camera to Scintillator Angle

The dependence of observed scintillator intensity on the angle to the camera was previously shown to be pseudo-Lambertian, maintaining a uniform response until a threshold angle of ~50 degrees.[13] In this work, the scintillator response was likewise characterized. Angle between the array and camera was defined as the difference between the vector from the isocenter to the known camera position and the normal vector from the scintillator face. To systematically vary the angle and evaluate the emission, the array was affixed flat to a 5 cm thick block of solid water and placed at the isocenter. The emission was characterized in 5 degree increments from 0 degrees (scintillator normal vector facing camera) to 75 degrees (clockwise rotation) by raising one side of the solid water until the surface registered at the correct angular increment while maintaining the center of the array at the isocenter. To remove the gantry angle impact, the gantry was rotated by the same angle to retain the source to surface distance and relative angle to the scintillator. 500 MU of a 15×15 cm² 6 MV field was delivered at each angular increment and the scintillation intensity from the array elements was averaged.

## 2.5 Water Equivalent Thickness

Water equivalent thickness was determined by measuring a Tissue Maximum Ratio (TMR) curve with and without having the array in the beam line. Depth-dependent measurements were obtained by stacking solid water (SW) blocks of varying thicknesses above an IBA RAZOR diode detector which was embedded in a 20 mm SW block, positioned on a 50 mm SW base. The source-to-detector distance was set to 100 cm with a 10×10 cm² field size. The measurement point was the inner face of the proximal electrode. For each measurement, 100 MU was delivered using 6 MV photon beams at 600 MU/min. Each depth was measured 4-5 times and averaged. TMR values for the linear accelerator were calculated by normalizing chamber readings to the maximum value. The scintillator array was then placed on the phantom surface above the chamber and irradiated under identical conditions. The measured charge was converted to TMR values, and the water-equivalent depth was determined from the established TMR-depth curve.

## 2.6 Field Edge Comparison

A field edge comparison was done to evaluate the accuracy of dose map interpolation over high dose gradient regions. The array was placed over a flat 5 cm block of solid water oriented at the isocenter. A sheet of gafchromic film was placed under the array and taped on the sides to remain flat. A 5×20 cm² 6 MV field was used to deliver 500 MU such that half of the array was out of the field light, and scintillator emission was collected using the Cherenkov camera. The scintillation profile was processed as outlined in section 2.3 and aligned to the film dose profile along the field edge. The resulting dose map was compared against the film profile using a 3%/3mm gamma analysis to highlight the extent of agreement over a gradient based on clinically relevant dose agreement constraints. The relative Y-axis profiles were likewise compared to observe the impact of camera blurring on edge identification.

## 2.7 Phantom Testing and Reproducibility

A single field delivery onto an anthropomorphic chest phantom was used to demonstrate clinical feasibility and evaluate the reproducibility of the imaging technique and downstream processing. 500 MU at 6MV of a tangent anteroposterior field was selected because contralateral breast dose monitoring represents a key clinical application with difficult to quantify unwanted dose and significant dose gradients. The scintillator array was positioned at the base of the contralateral breast surface and continued onto the sternum, placing the edge of the field close to the midline of the array. Five separate but consecutive deliveries were performed without changing the setup, with each processed through the complete analysis pipeline. Subsequent measurements were compared to the first delivery to evaluate the robustness and stability of the reconstruction algorithm and evaluate the reproducibility in the surface dose reconstruction based on the inherent uncertainties in the processing pipeline.

## 3. Results

### 3.1 Beam Response and Water Equivalent Thickness

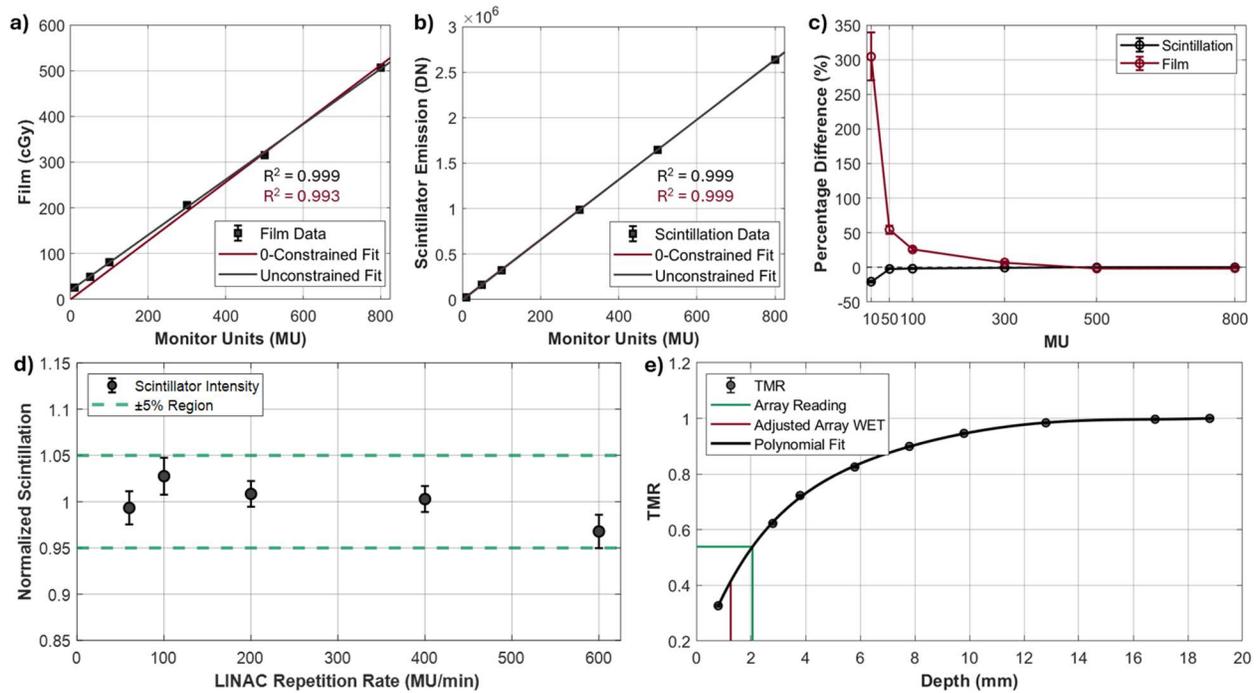

**Figure 4: a)** Film linearity with MU is shown with linear fits both constrained through zero and unconstrained at the y-intercept. **b)** Average scintillation emission from the array versus MU is similarly fit using both constrained and unconstrained linear models. **c)** Residuals for film and scintillation are compared, revealing expected deviations at low MU values. **d)** Scintillation response with pulse repetition rate is shown to stay within 5% for all available rates on the TrueBeam LINAC. **e)** WET evaluation using an experimentally derived TMR curve indicates an array thickness of 1.21 mm.

The scintillator array was tested under varying delivery conditions to characterize its response properties and physical impact on beam delivery. To establish a reliable ground truth comparison to dose, film response was measured at increasing MU values and fit to two linear functions with and without constraining the intercept to zero (Figure 4a), since an ideal dosimeter is expected to exhibit linear response through zero. Scintillation response was similarly characterized using both fitting approaches, shown in Figure 4b. Residuals from the zero-constrained fit revealed expected film deviation from linearity at low MU values compared to scintillation.[25,26] Both detectors showed minimal residuals above 300 MU, leading to the selection of 500 MU for scintillation-to-dose conversion (Figure 4c). Scintillation response under all available repetition rates on the TrueBeam LINAC was tested, with intensity varying by less than 5% across all rates, as shown in Figure 4d. For water equivalent thickness determination, TMR was measured across multiple depths and fit with a polynomial function. The scintillator array demonstrated a WET of 2.01 mm when placed in the beam. After adjusting for the 0.8 mm measurement depth, the corrected array WET was 1.21 mm, closely matching the physical array thickness of 1.1 mm.

### 3.2 Angular Dependence

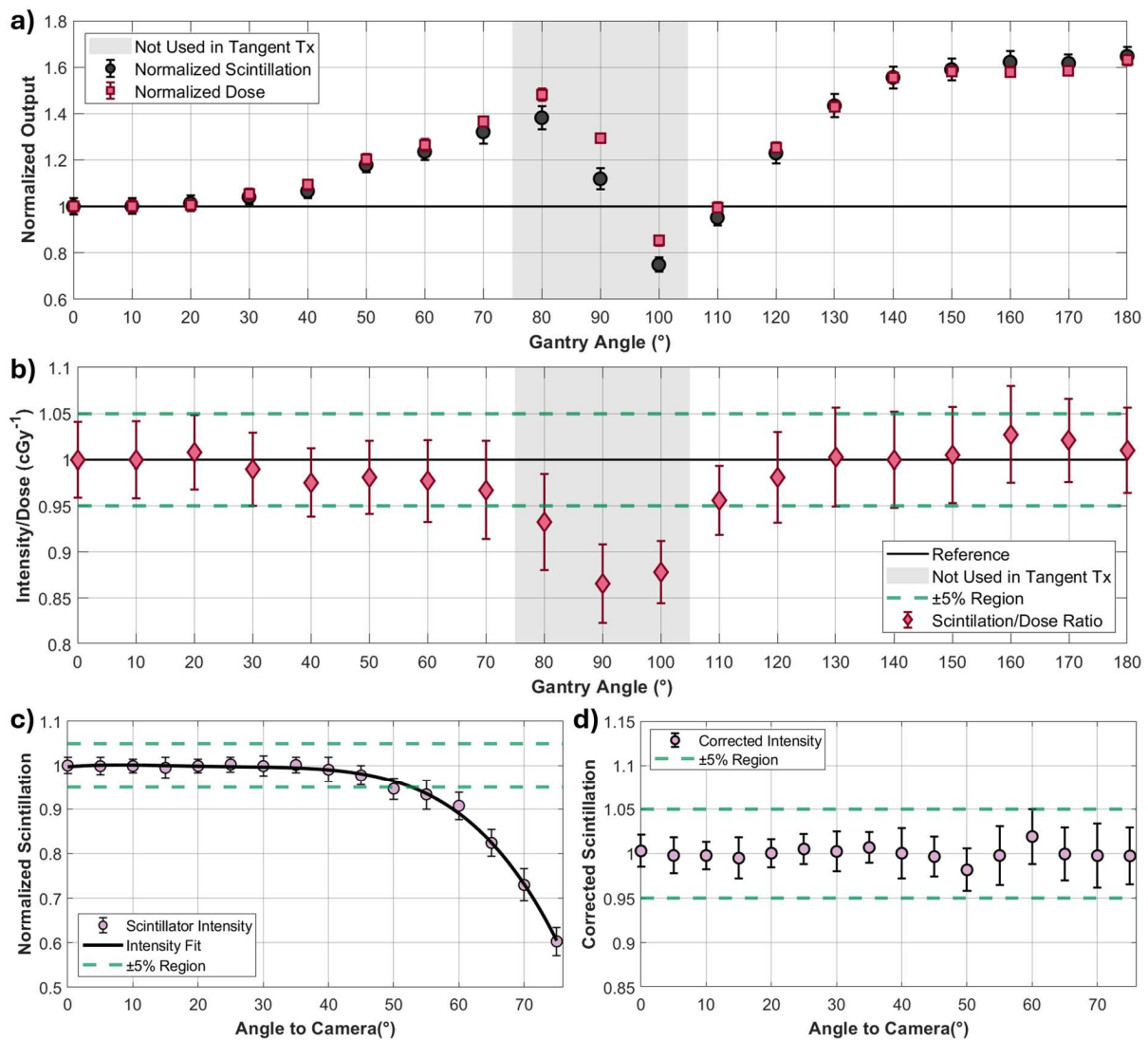

**Figure 5:** Scintillator response variability with gantry angle is shown in (a) over 0-180 degrees alongside film response at corresponding angles. Dose-corrected scintillation response in (b) remains within 5% of the 0-degree emission in the region of interest for tangent treatments. Scintillation emission with camera angle is presented in (c) with the corresponding corrected emission in (d) based on the intensity fit, which maintains values within 5% of the mean across the full 0-75 degree range.

Tangent breast treatments typically use gantry angles ranging from 0 to 75 degrees, defining the clinically relevant range for dosimetric evaluation.[27] The scintillator array conforms to patient surface geometry and deforms with respiratory motion and general patient movement, resulting in a highly variable distribution of viewing angles relative to the camera and irradiation angles relative to the gantry throughout treatment. Scintillation intensity was observed in Figure 5a to deviate up to 60% over the full angular range, but was closely matched by dose reported by film. Dose-normalized scintillation output, shown in Figure 5b, deviated less than 5% across the target range of gantry angles, suggesting angle-dependent corrections with respect to gantry position was not necessary for the target range. In contrast, angular deviation with camera angle was observed to be pseudo-Lambertian, remaining within 5% until 45 degrees, and decreasing in emission exponentially until the end of the tested range at 75 degrees. Camera viewing angle corrections were therefore derived based on the intensity fit shown in Figure 5c, and applied on a per-element basis as outlined in Figure 2. Following correction, values remained within 5% of the mean response across the entire 0-75 degree range tested.

### 3.3 Field Edge Verification

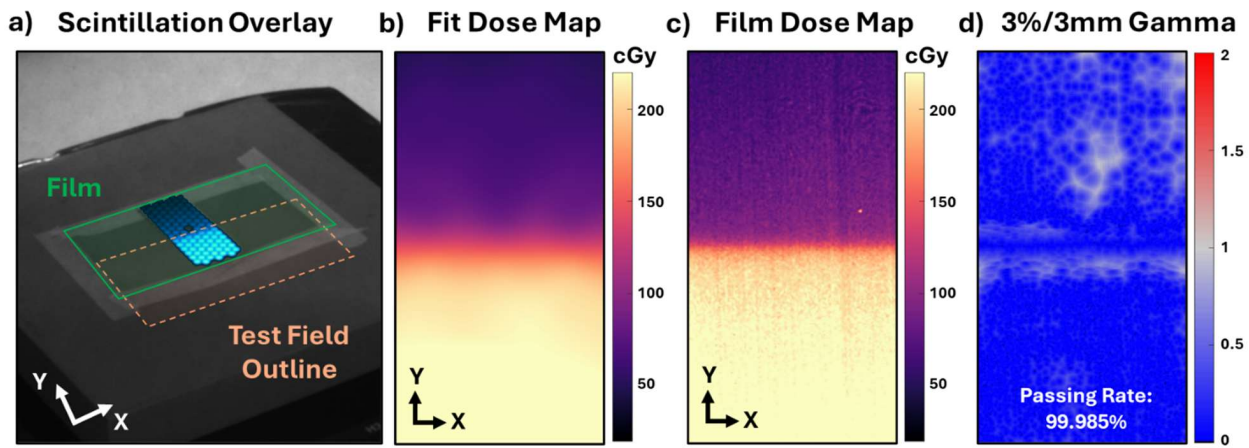

**Figure 6: (a)** Scintillation output overlay onto background is shown with placement outlines of film and the edge test field. En face comparison of fit dose map **(b)** and film dose map **(c)** was done using a 3%/3mm gamma analysis and is shown **(d)** with a 99.98% passing rate.

Field edge response was evaluated to assess the scintillator array performance in high dose gradient regions. Shown in Figure 6a, a delivery onto a flat phantom was done with concurrent film measurement to establish ground truth comparison. To test system accuracy in a demanding orientation, the array was lined up with field edge as opposed to having a staggered orientation of elements which would be commonly seen in a clinical application over complex anatomy. This alignment ensured that the steep dose gradient of 25.6 cGy/mm (calculated from film dose map) fell partially between scintillator elements, testing the interpolation robustness. Gamma analysis done at 3%/3mm between interpolated scintillator dose maps (Figure 6b) and film measurements (Figure 6c) is shown in Figure 6d and yielded a 99.98% passing rate, indicating high agreement.

**3.4 Phantom Testing and Reproducibility**

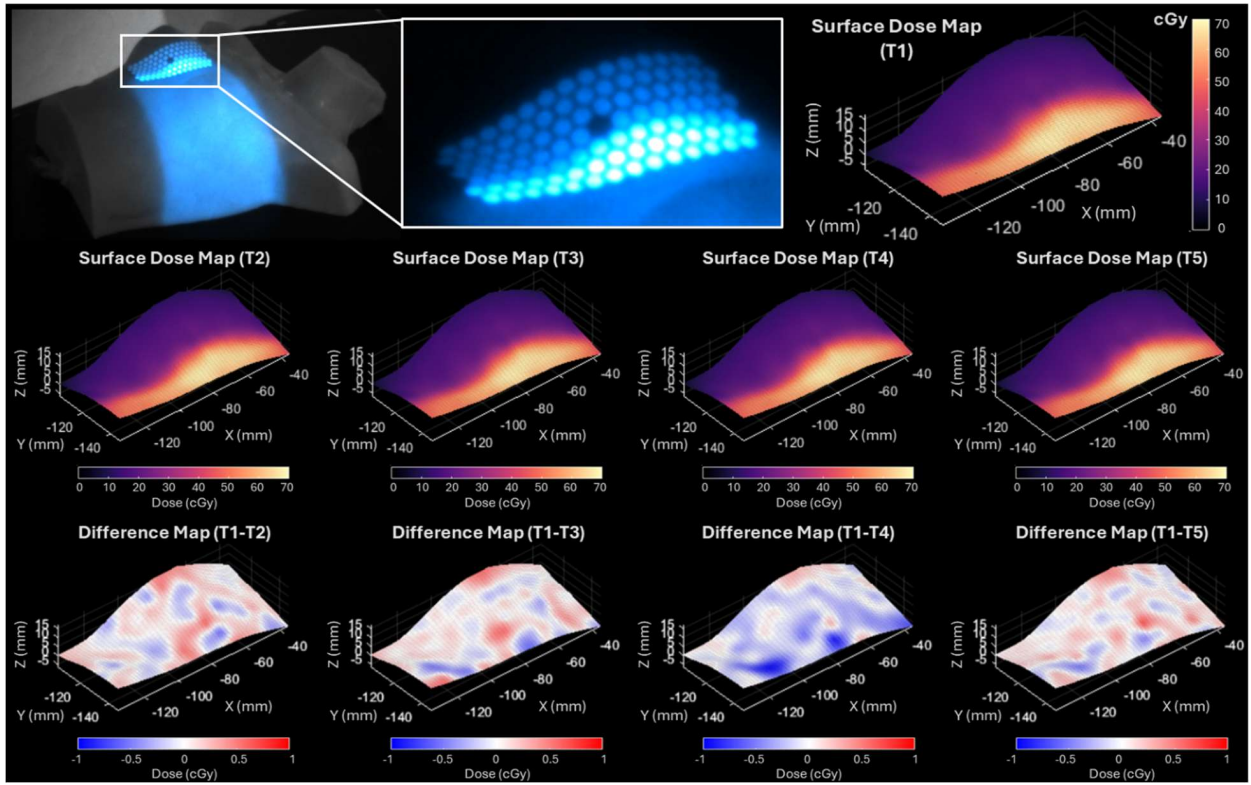

**Figure 7:** Complete testing of the proposed system is shown as a contralateral breast surface dose monitoring application during a right anterior-oblique tangent field delivery. Raw scintillator output is highlighted. Cumulative surface dose map is shown in 3D for each of the 5 trial deliveries (T1 – T5). Difference maps of trial deliveries 2 through 5 (T2 – T5) are shown compared to trial delivery 1 (T1) highlighting variability in processing between deliveries.

A complete implementation of the scintillator array imaging system was tested on an anthropomorphic chest phantom by delivering a right anterior-oblique tangent field and monitoring the contralateral breast surface. The stereovision pipeline was used to localize the array on the phantom surface, link the position information to the scintillator output map, and leverage spline interpolation to generate a continuous surface. Normal vectors at each element yielded appropriate scaling factors for camera angle correction. The scintillation-to-dose relationship was established from the 500 MU film comparison and the cumulative surface dose map is shown in 3D for the delivered field. Five consecutive deliveries (T1-T5) were performed with complete processing pipeline repetition to evaluate system reproducibility. Subsequent deliveries (T2-T5) showed deviations within a range of ±1 cGy, and an average magnitude of 0.14 (0.11) cGy compared to the initial delivery (T1). These variations fell well within the expected LINAC output variability of ±3% (daily QA), demonstrating that the system could have sufficiently robust performance for beam quantification of surface dose in this application.

## 4. Discussion

This study presents the characterization and translation of broad area conformable scintillator array imaging as a surface dosimetry technique for conventional photon external beam radiotherapy. Prior work involving scintillator array dosimetry focused on UHDR PBS proton therapy and delivery verification[11], which involved known spot profiles and high-speed delivery. In translating to conventional photon therapy, the scintillator behavior is non-trivial given the build-up effects of megavoltage photon beams, temporal delivery dynamics, and large area dose gradients. The characterization presented herein addresses these differences and evaluates the feasibility of scintillator array dosimetry in conventional photon EBRT.

The system integrates a modular scintillator array design with stereovision-based 3D localization and real-time optical imaging. The array consists of 101 hexagonal elements over a 10×4 cm² area which function as independent dosimeters with internally spatially resolved dose response while maintaining collective conformability to target surfaces with conformality to curvatures of up to 5mm radius. Their response can be interpolated to derive a single continuous dose map. The fragmented design enables both precise spatial localization and element-specific angular correction which is unavailable in existing dosimetry systems. Traditional alternatives have significant limitations as TLDs, OSLDs, and MOSFETs are point dosimeters and lack spatial correlation to the target while radiochromic film offers 2D coverage but cannot conform to curved surfaces, requires delayed readout, and exhibits increasingly nonlinear response at low doses. Continuous flexible scintillation emitters address the temporal requirement, but cannot provide the discretized angular corrections necessary for dose conversion from optical emission.

The locally rigid array design maintained measurement accuracy across varying beam and viewing angles. Each hexagonal element preserves consistent emission characteristics during array deformation, enabling application of viewing angle corrections derived from stereovision surface reconstruction. Angular characterization revealed dose-normalized intensity variations within 5% for gantry angles from 0-75 degrees, encompassing the clinically relevant range for tangent treatments.

Camera angle corrections successfully compensated for the pseudo-Lambertian emission profile, maintaining response within 5% across the tested range. Angular response degradation beyond 75 degrees limits applicability for certain beam configurations, but this primarily resulted from challenges obtaining reliable reference measurements at extreme angles. The discrete elements represent a trade-off between angular correction capability and continuous surface coverage, with dose information between elements dependent on interpolation accuracy. Nevertheless, demonstrated performance at field edges indicates adequate capture of clinically relevant gradients.

Functionally, the system achieved minimal beam perturbation with a water equivalent thickness of 1.21 mm, suggesting minimal impact on surface dose deposition. High dose linearity ($R^2 = 0.999$) was maintained across the full tested range of 10-800 MU, compared to film which exhibited an expected non-linear response at low MU. This film behavior was anticipated as the system used was optimized for clinical in-field dosimetry with calibration prioritized for the >2Gy region. Slight under response was observed in the scintillator at the 10 MU delivery, which was attributed to overcorrection during background subtraction due to the noise-dependent fluctuations. Scintillation response remained consistent within 5% across all LINAC repetition rates (60-600 MU/min), confirming stability under varying delivery conditions. Inter-delivery reproducibility average deviation of 0.14 (0.11) cGy over a 0-70cGy range demonstrated robust system performance, with output variation well expected LINAC output fluctuation. As a field edge dosimetry technique, the scintillator array dose map achieved a 99.98% gamma pass rate (3%/3mm) compared to film. Observed deviations between scintillation and film dose maps in the penumbral region were expected given the clinical Cherenkov system's f/2 aperture, which caused focus decline farther from the camera and effective blurring evident in Figure 5 compared to Figure 6 which was taken at a more clinically appropriate field depth.

Despite these promising results, several limitations exist. Optical detection restricts measurements to camera-visible surfaces, potentially limiting coverage for certain anatomical sites. While this is an issue with a single camera view, a multiple camera configuration could address such angular limitations and visibility constraints. Our current validation focused on static field deliveries, and an extension to VMAT would be necessary for this system to be fully clinically viable, which would require time resolved angular corrections during irradiation in the 80-110 degree range. Clinical translation will also require prospective patient studies, particularly for treatments where surface dosimetry is of interest such as gradient dosimetry, field matching, or treatments with avoidance areas. The 3D localization from stereovision imaging in direct comparison with treatment planning system predictions should also be investigated as this would enable direct dose comparison and act as a secondary intra-treatment quality assurance tool. Future work should also expand the scintillator response testing to the full available clinical energy range since, although no energy dependence is expected given the negligible Cherenkov contribution to the optical signal, a quantitative analysis would be valuable prior to clinical translation. Additionally, the added 1.21mm WET imposed by the array introduces a <0.1% shift in the tissue phantom ratio (TPR) calculation at 4-5cm depth which may be considered to have negligible impact on the dose distribution. However, by optimizing the thickness of the support material, the array WET could be brought to below 1mm, making it more suitable for *in vivo* applications. Finally, observed deviations due to inter-element interpolation accuracy could also be addressed in future work through informed interpolation.

## 5. Conclusion:

As radiotherapy advances toward increasingly conformal and dynamic techniques, in vivo dosimetry should become as informative as possible to allow for efficient and effective verification of delivered dose. This work introduced a wide-area conformal scintillation imaging system for photon EBRT capable of time resolved, in vivo surface dosimetry with minimal workflow impact. The modular design and stereovision localization introduced with this system presents the unique benefits of conformality over curved surfaces, independent localization in room coordinates, and 2D dose derivation which is not achievable with standard point dosimetry. These features also allow the system to resolve steep dose gradients independent of spatial positioning which offers additional utility in field edge and out of field dose monitoring. Ongoing work will further test this technology in clinical use, in a variety of relevant anatomical sites to evaluate the range of its utility. Future technological development may focus on expanding compatibility with dynamic treatment techniques and improving the spatial interpolation between measurement points to maximize clinical benefit.

5. Olaciregui-Ruiz I, Beddar S, Greer P, Jornet N, McCurdy B, Paiva-Fonseca G, Mijnheer B, Verhaegen F. In vivo dosimetry in external beam photon radiotherapy: Requirements and future directions for research, development, and clinical practice. Phys Imaging Radiat Oncol. 2020 Jul;15:108–116.

6. Verhaegen F, Fonseca GP, Johansen JG, Beaulieu L, Beddar S, Greer P, Jornet N, Kertzscher G, McCurdy B, Smith RL, Mijnheer B, Olaciregui-Ruiz I, Tanderup K. Future directions of in vivo dosimetry for external beam radiotherapy and brachytherapy. Phys Imaging Radiat Oncol. 2020 Oct;16:18–19. PMCID: PMC7807862

7. MacDougall ND, Graveling M, Hansen VN, Brownsword K, Morgan A. In vivo dosimetry in UK external beam radiotherapy: current and future usage. Br J Radiol. 2017 Apr;90(1072):20160915. PMCID: PMC5605079

8. Nailon WH, Welsh D, McDonald K, Burns D, Forsyth J, Cooke G, Cutanda F, Carruthers LJ, McLaren DB, Puxeu Vaqué J, Kehoe T, Andiappa S. EPID-based in vivo dosimetry using Dosimetry Check™: Overview and clinical experience in a 5-yr study including breast, lung, prostate, and head and neck cancer patients. J Appl Clin Med Phys. 2019 Jan;20(1):6–16. PMCID: PMC6333145

9. Yusof FH, Ung NM, Wong JHD, Jong WL, Ath V, Phua VCE, Heng SP, Ng KH. On the Use of Optically Stimulated Luminescent Dosimeter for Surface Dose Measurement during Radiotherapy. PloS One. 2015;10(6):e0128544. PMCID: PMC4459977

10. Yao T, Luthjens LH, Gasparini A, Warman JM. A study of four radiochromic films currently used for (2D) radiation dosimetry. Radiat Phys Chem. 2017 Apr;133:37–44.

11. Vasyltsiv R, Harms J, Clark M, Gladstone DJ, Pogue BW, Zhang R, Bruza P. Design and characterization of a novel scintillator array for UHDR PBS proton therapy surface dosimetry. Med Phys. 2025 May 31; PMID: 40450336

12. Kanouta E, Bruza P, Johansen JG, Kristensen L, Sørensen BS, Poulsen PR. Two-dimensional time-resolved scintillating sheet monitoring of proton pencil beam scanning FLASH mouse irradiations. Med Phys. 2024 Jul;51(7):5119–5129.

13. Decker SM, Bruza P, Zhang R, Williams BB, Jarvis LA, Pogue BW, Gladstone DJ. Technical note: Visual, rapid, scintillation point dosimetry for in vivo MV photon beam radiotherapy treatments. Med Phys. 2024 Aug;51(8):5754–5763.

14. Clark M, Harms J, Vasyltsiv R, Sloop A, Kozelka J, Simon B, Zhang R, Gladstone D, Bruza P. Quantitative, real-time scintillation imaging for experimental comparison of different dose and dose rate estimations in UHDR proton pencil beams. Med Phys. 2024 Sep;51(9):6402–6411.

15. Jia M, Yang Y, Wu Y, Li X, Xing L, Wang L. Deep learning-augmented radioluminescence imaging for radiotherapy dose verification. Med Phys. 2021 Nov;48(11):6820–6831. PMID: 34523131

16. Jia M, Kim TJ, Yang Y, Xing L, Jean PD, Grafil E, Jenkins CH, Fahimian BP. Automated multi-parameter high-dose-rate brachytherapy quality assurance via radioluminescence imaging. Phys Med Biol. 2020 Nov 17;65(22):225005. PMCID: PMC7755302
18